\documentclass[reprint, prl, twocolumn, superscriptaddress]{revtex4-1}
\usepackage{amssymb, graphics, epsfig, color, ulem, float,graphicx}
\usepackage{amsmath,amstext,bm,verbatim}
\usepackage{graphicx}
\usepackage{epstopdf}

\begin{document}

\title{Salt-dependent rheology and surface tension of protein condensates using optical traps}
\author{Louise M. Jawerth}
\affiliation{Max-Planck-Institut f\"ur Zellbiologie und Genetik, Pfotenhauerstra{\ss}e 108, 01307 Dresden, Germany}
\affiliation{Max-Planck-Institut f\"ur Physik komplexer Systeme, N\"othnitzerstra{\ss}e 38, 01187 Dresden, Germany}
\author{Mahdiye Ijavi}
\affiliation{Max-Planck-Institut f\"ur Zellbiologie und Genetik, Pfotenhauerstra{\ss}e 108, 01307 Dresden, Germany}
\author{Martine Ruer}
\affiliation{Max-Planck-Institut f\"ur Zellbiologie und Genetik, Pfotenhauerstra{\ss}e 108, 01307 Dresden, Germany}
\author{Shambaditya Saha}
\affiliation{Max-Planck-Institut f\"ur Zellbiologie und Genetik, Pfotenhauerstra{\ss}e 108, 01307 Dresden, Germany}
\author{Marcus Jahnel}
\affiliation{Max-Planck-Institut f\"ur Zellbiologie und Genetik, Pfotenhauerstra{\ss}e 108, 01307 Dresden, Germany}
\affiliation{Biotechnology Center, Technische Universit\"at Dresden,  Tatzberg 47-49, 01307 Dresden, Germany}
\author{Anthony A. Hyman}
\affiliation{Max-Planck-Institut f\"ur Zellbiologie und Genetik, Pfotenhauerstra{\ss}e 108, 01307 Dresden, Germany}
\author{Frank J\"ulicher}
\email{Corresponding author: julicher@pks.mpg.de}
\affiliation{Max-Planck-Institut f\"ur Physik komplexer Systeme, N\"othnitzerstra{\ss}e 38, 01187 Dresden, Germany}
\author{Elisabeth Fischer-Friedrich}
\email{Corresponding author: elisabeth.fischer-friedrich@tu-dresden.de}
\affiliation{Max-Planck-Institut f\"ur Zellbiologie und Genetik, Pfotenhauerstra{\ss}e 108, 01307 Dresden, Germany}
\affiliation{Max-Planck-Institut f\"ur Physik komplexer Systeme, N\"othnitzerstra{\ss}e 38, 01187 Dresden, Germany}
\affiliation{Biotechnology Center, Technische Universit\"at Dresden,  Tatzberg 47-49, 01307 Dresden, Germany}

\date{\today}

\begin{abstract}
An increasing number of proteins with intrinsically disordered domains have been shown to phase separate in buffer to form liquid-like phases. These protein condensates serve as simple models for the investigation of the more complex membrane-less organelles in cells. 
To understand the function of such proteins in cells, the material properties of the condensates they form are important. However, these material properties are not well understood.
Here, we develop a novel method based on optical traps to study the frequency-dependent rheology and the surface tension of PGL-3 condensates as a function of salt concentration. 
We find that PGL-3 droplets are predominantly viscous but also exhibit elastic properties.
As the  salt concentration is reduced, their elastic modulus, viscosity and surface tension increase. 
Our findings show that salt concentration  has a strong influence on the rheology and dynamics of protein condensates suggesting an important role of electrostatic interactions for their material properties.
\end{abstract}
\maketitle

\begin{figure}
\centering
\includegraphics[width=7.5 cm]{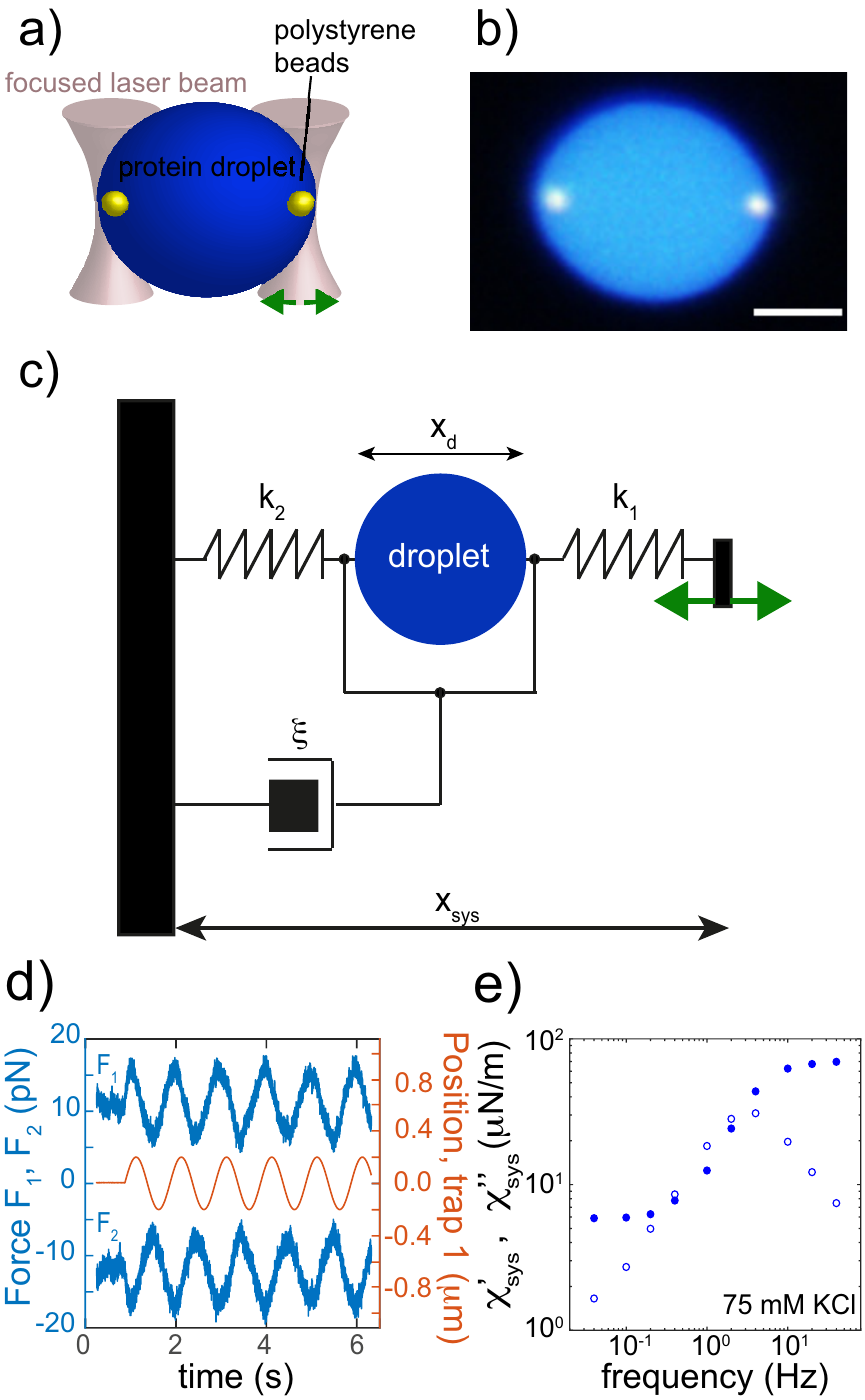}
\caption{ \label{fig:Fig1}
Microrheology setup based on a dual optical trap.
a) Schematic.
b) Fluorescent micrograph of protein droplet during the measurement. Scale bar: $5\,\mu$m.
c) Mechanical circuit equivalent of measurement setup.
d) Exemplary experimental output during oscillatory droplet deformation: position of optical trap 1 (red) and forces associated to optical traps 1 and 2 (blue)  at frequency 1~Hz and 75~mM KCl.
e) Exemplary system spring constants $\chi^\prime_{sys}$ ($\bullet$), $\chi^{\prime\prime}_{sys}$ ($\circ$) from the measurement of a protein droplet with a diameter of $~12.2\,\mu$m at a salt concentration of 75~mM KCl. 
}
\end{figure}

{\it Introduction.} 
A fundamental question of biology is to understand the spatial organization of cells into compartments of distinct chemical composition and activity.  
Many cellular compartments are bounded by membranes. However, cells also possess compartments that are not bounded by membranes. Examples are germline granules, cajal bodies, nucleoli and stress granules. It has been shown in recent years that many membrane-less compartments are protein condensates forming soft materials that coexist with the cytoplasm or nucleoplasm and often exhibit liquid-like properties  \cite{hyma14, bran09, bran11, gilk04, pate15, bran09, bran11, saha16, lin15, bana16,  mura15, nott15}.\\
Liquid-like compartments typically contain one or two scaffold proteins which are required for compartment formation and to which RNA and other proteins co-localize \cite{hyma14, gilk04, aoki16}. Many scaffold proteins phase separate {\it in vitro} when introduced into physiological buffer resulting in the condensation of liquid-like droplets. A well-studied example of a phase-separating scaffold protein is PGL-3 - a major component of P-granules, the germline granules of the nematode worm {\it Caenorhabditis elegans} \cite{bran09, elba15,saha16, aoki16}. This and other protein droplets serve as models for the soft materials that form the more complex  liquid-like compartments in cells \cite{moll15, saha16, lin15, nott15, elba15, pate15}.  The material properties of protein droplets are important to understand the biological function of protein condensates in cells - for example when they serve as biochemical reaction centers, which requires rapid diffusion of components, or when they serve as structural elements, which may require some degree of elasticity. However, the material properties and rheology have not yet been fully characterized. Active microrheological methods that could provide such information for micron-sized droplets are currently lacking. \\
The saturation concentration of protein phase separation {\it in vitro} depends on buffer conditions such as salt concentration, pH or the presence of RNA\cite{moll15, saha16, lin15, nott15, elba15, bran15}. The salt and pH dependence of the saturation concentration suggests a role of charge and electrostatic interactions in protein phase separation.  \\
In this letter, we develop a novel active microrheology method to determine the frequency-dependent rheology of micron-sized PGL-3 droplets as a function of salt concentration using optical traps. Protein droplets are deformed via two opposing adhered beads trapped by optical traps (Fig.~\ref{fig:Fig1}a,b). Surface tension $\gamma$ and the complex shear modulus $G^{\ast}(\omega) = G^\prime(\omega)  + iG^{\prime\prime}(\omega) $ that characterize the droplet material properties can be obtained using a theoretical analysis of force balances associated with droplet deformation in the optical trap setup. Here $G^\prime$ and $G^{\prime\prime}$ denote the storage and loss moduli, respectively.\\
{\it Force balance of droplet deformation.}
The microrheology setup based on a dual optical trap is shown in Fig.~\ref{fig:Fig1}a,b.
The droplet is deformed by moving one trap. The resulting force balance involves forces $F_1$ and $F_2$ exerted on the droplet by the traps with stiffnesses $k_1$ and $k_2$, as well as viscous drag exerted on the droplet by the fluid with friction coefficient $\xi$ (Fig.~\ref{fig:Fig1}c).
In the over-damped regime, force balance requires 
\begin{eqnarray}
F_1+F_2&= \xi v   \label{eq:ForcBal2}\quad,
\end{eqnarray}
where $v$ is the droplet velocity.
The difference of $F_1$ and $F_2$ is balanced by forces exerted by the deformed droplet. For time periodic forces $F_{1,2} = \tilde F_{1,2} \exp(i\omega t) + c.c.$ and droplet diameter $ x_{d}= x_{d,0}+ \tilde x_{d} \exp(i\omega t)+ c.c.$ with angular frequency $\omega$, we have
\begin{eqnarray}
(\tilde F_1 - \tilde F_2)/2&= \chi^\ast(\omega) \tilde x_{d} \label{eq:ForcBal1} \quad.
\end{eqnarray}
Here, the tilde denotes the complex amplitude of the Fourier mode at angular frequency $\omega$ and $c.c.$ refers to the complex conjugate. In Eq.~\eqref{eq:ForcBal1}, $\chi^\ast(\omega)=\chi^\prime(\omega)+i\chi^{\prime\prime}(\omega) $ characterizes the complex frequency-dependent spring constant of the droplet response, where $\chi^\prime(\omega)$ and $\chi^{\prime\prime}(\omega)$ denote the real and imaginary part of the complex spring constant, respectively. For an oscillatory trap movement $\tilde v=i \omega (\tilde x_{d}/2 +\tilde F_2/k_2)$.
The complex spring constant of the entire system (droplet in series with two optical traps) is 
\begin{equation}
\chi^\ast_{sys}(\omega)=(\tilde F_1-\tilde F_2)/(2 \tilde x_{sys}) \label{eq:SysStiff}
\end{equation}
where $ x_{sys}$ is the distance between the trap centers (Fig.~\ref{fig:Fig1}c). The distance $ x_{sys}$ can differ from the droplet diameter $x_d$ because the beads can move out of the trap centers. The complex spring constant of the droplet can be related to the complex spring constant of the system using
\begin{equation}
x_{sys}=(F_1/k_1 - F_2/k_2)+ x_{d} \label{eq:xsys_xdro}\quad.
\end{equation} 
We find
\begin{equation}
\chi^\ast=\frac{\chi^\ast_{sys} (4 k_1 k_2 + i  \xi \omega (k_1+k_2))}{
 2 k_1 (2 k_2 + i \xi \omega)-4\chi^\ast_{sys} (k_1 + k_2 + i \xi \omega) } \label{eq:DroStiff}\quad.
 \end{equation}

\begin{figure*}
\centering
\includegraphics[width=\textwidth]{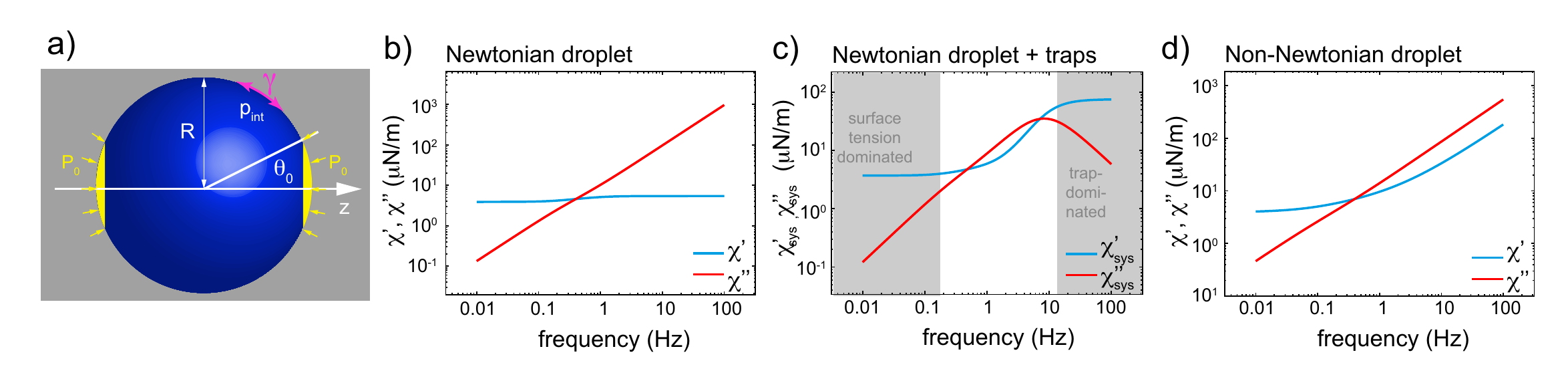}
\caption{ \label{fig:Fig2}
Spring constants for a Newtonian and non-Newtonian-droplet subject to polar forcing.
a) Droplets are subject to an oscillatory pressure $P_0$ applied in a dome-shaped region at the poles.
b) Complex spring constant $\chi^\ast$ of  a Newtonian droplet with surface tension.
c) Complex system spring constant of a Newtonian droplet in series with two identical traps.  For low frequencies, the spring constant is dominated by surface tension (grey zone, l.h.s.). For high frequencies, the two traps dominate the spring constant and $\chi^\prime_{sys}$ approaches $k/2$ (grey zone, r.h.s.). Outside of this trap-dominated regime, the loss modulus is governed by the viscous properties of the droplet material. 
d) Spring constant of  a non-Newtonin droplet formed by a power law fluid, i.e. $G^\ast(\omega)= A\pi\omega^\beta  (\csc(\pi \beta/2)+i \sec(\pi \beta/2))/2$, and with surface tension. 
Parameters are $\theta_0=0.1, R=5\,\mu{\rm m}, \gamma=3\,\mu{\rm N/m}, \eta=0.5\,{\rm Pa s}, A=0.2\,{\rm Pa}, \beta=0.8$, $k_{1}=k_2=150\,$pN/$\mu$m and $\xi=6\pi \eta_b R$, where $\eta_b=10^{-3} {\rm Pa}\,$s.
}
\end{figure*}

{\it Droplet mechanics in the optical trap.}
To determine the protein droplet's surface tension $\gamma$ as well as the complex frequency dependent shear modulus $G^{\ast}(\omega)$ in the optical trap setup, we need to relate these material properties to the complex spring constant $\chi^\ast$ defined in Eq.~\eqref{eq:ForcBal1}.
The mechanical droplet stress satisfies the force balance
\begin{equation}
\partial_i\sigma_{ij}=0 \label{eq:divsig} \quad.
\end{equation}
For small deformations in the low Reynold's number regime, the constitutive material equation of an incompressible droplet material can be written as
\begin{equation}
\tilde \sigma_{ij} = - \tilde p \cdot\delta_{ij} +G^\ast(\omega)  \left[\partial_i \tilde u_j +\partial_j \tilde u_i\right]\quad,\label{eq:stressansatz}
\end{equation}
where $p$ denotes pressure and $u_i$ is the displacement vector with components $i,j=x,y,z$.
Stress boundary conditions on the surface of a spherical droplet of radius $R$ read in spherical coordinates $r$, $\theta$, $\phi$
\begin{eqnarray}
-p_{ext}- \sigma_{rr}(R,\theta) & = 2 \gamma H \label{eq:bound1}\\
\sigma_{r,\theta}(R,\theta) &=0\quad. \label{eq:bound2}
\end{eqnarray}
Here, $p_{ext}$ is the local normal force per area exerted on the droplet surface  and $H$ denotes the mean curvature of the weakly deformed droplet surface.  
Forces mediated by adherent beads of radius $r_{bead}$ are captured by a time periodic pressure distribution with complex amplitude $ \tilde p_{ext}=P_0 (\Theta(\theta- \pi+\theta_0)+\Theta(\theta_0-\theta))$ on the droplet surface (Fig.~\ref{fig:Fig2}a), where $\Theta$ denotes the Heaviside function and $\theta_0=r_{bead}/R$.

Using Eqn.~(\ref{eq:divsig}-\ref{eq:bound2}), we calculate the time-dependent deformation field $u_r(r,\theta, t)$ of droplet material using a stream function approach and an expansion in spherical harmonics, see Supplement. The time dependent droplet diameter can be expressed as $x_d(t)=2(R+u_r(R,0,t))$ and the difference of optical trap forces is given by $F_1(t)-F_2(t) = \oint dA\, p_{ext}$. We then obtain the complex spring constant of the droplet response $\chi^\ast$ as a function of $G^{\ast}$ and $\gamma$ using Eq.~\eqref{eq:ForcBal1}, see Supplement. This droplet spring constant is related to the spring constant of the entire optical trap system through Eq.~\eqref{eq:DroStiff}. 

\begin{figure}
\centering
\includegraphics[width=7.8cm]{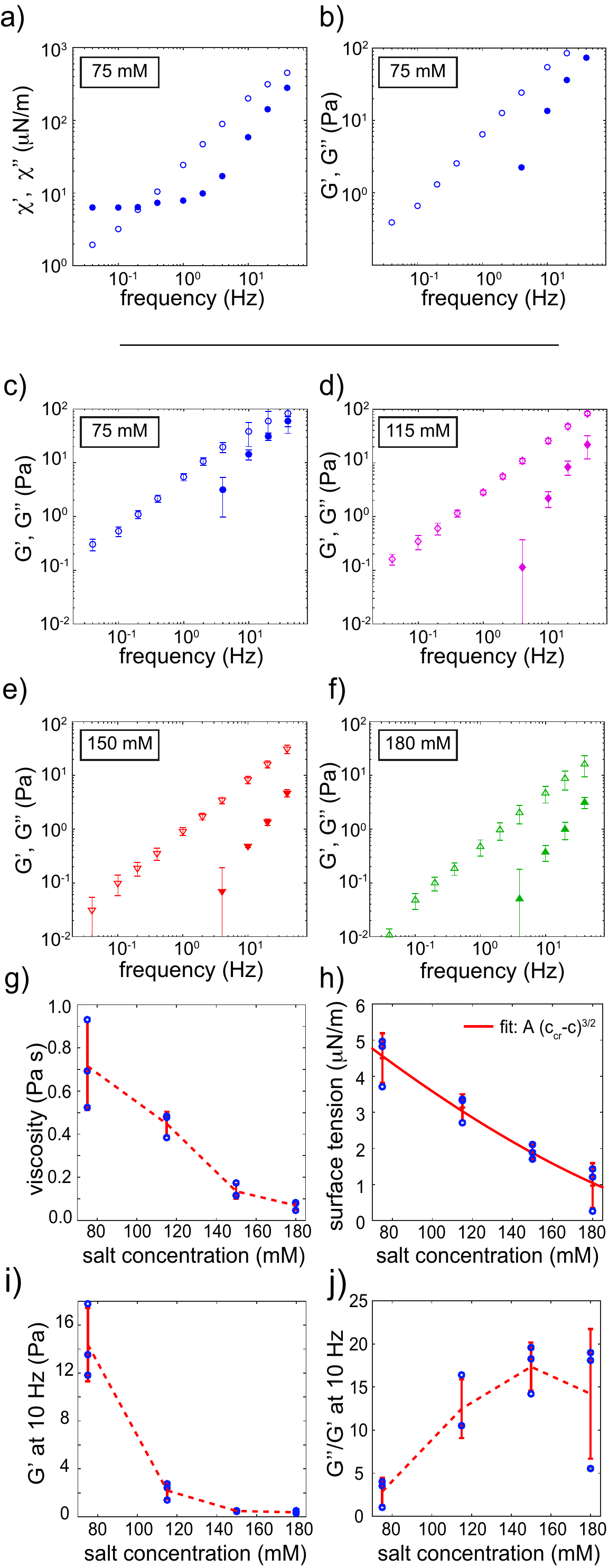}
\caption{ \label{fig:Fig3}
Salt dependent material properties of PGL-3 droplets. a-b) Spring constants $\chi^\prime$ ($\bullet$), $\chi^{\prime\prime}$ ($\circ$) and reconstructed moduli $G^\prime$ ($\bullet$), $G^{\prime\prime}$ ($\circ$) for one droplet. (Corresponding $\chi^\ast_{sys}$ shown in Fig.~\ref{fig:Fig1}e) c-f) Ensemble average of $G^\prime$ (closed symbols), $G^{\prime\prime}$ (open symbols) of the droplet material at salt concentrations of $75, 115, 150$ and $180\,$mM KCl.  Error bars indicate standard deviations (n=3). 
 g-j) Salt dependence of viscosity found using a linear fit to the loss modulus ($G^{\prime\prime}=\omega\eta$ where $\eta$ is the viscosity) (g), surface tension (h), storage modulus at $10\,$Hz (i) and loss tangent at $10\,$Hz (j). {\color{black} 
In panel h, we show a fit of the theoretically predicted decay of surface tension in dependence of salt concentration $c$ (solid red line) \cite{spru10b}, where  $c_{cr}=243$~mM is the fitted critical salt concentration.
 }
 }
\end{figure}

{\it Droplet spring constant in simple cases.}
We first discuss the droplet and system spring constants for a droplet consisting of a Newtonian fluid. In this case, the complex shear modulus $G^\ast = i \omega \eta$ where $\eta$ is the viscosity. 
Using the approach described above, we determine the droplet spring constant $\chi^\ast$ and the system spring constant $\chi_{sys}^\ast$. The droplet spring constant as a function of frequency is shown in Fig. ~\ref{fig:Fig2}b. The real part of the spring constant (blue) stems from surface tension. The imaginary part (red) is associated with viscosity. \\
The system spring constant for a Newtoninan droplet is shown in Fig. ~\ref{fig:Fig2}c. At low frequencies, the response of the system is dominated by the droplet surface tension, corresponding to a plateau in the real part (see gray low-frequency region Fig.~\ref{fig:Fig2}c).
The intermediate frequency range is dominated by droplet viscosity. At high frequencies, the system response is dominated by the optical traps (see gray high-frequency region Fig.~\ref{fig:Fig2}c). \\
We also consider a droplet composed of a non-Newtonian fluid with a power law rheology in both loss and storage moduli $G^\prime\sim G^{\prime\prime}\sim \omega^\beta$ with exponent $\beta$. The corresponding droplet spring constant reveals droplet surface tension at low frequencies and power-law rheology at high frequencies, Fig.~\ref{fig:Fig2}d.  
\\
{\it Salt dependent material properties of PGL-3 droplets.}
We performed measurements on PGL-3 protein droplets formed in buffer (25mM HEPES, 1mM DTT, pH 7.5) at four different salt concentrations (75 mM, 115 mM, 150 mM and 180 mM KCl).  Polystyrene beads of $1\mu$m diameter added to the buffer were brought into adhesive  contact with protein droplets {\color{black} (see Supplement). Radii of  droplets used were between $5\text{-}9\,\mu$m.}
Droplets were {\color{black}} deformed at  frequencies between $0.04$ and $40\,$Hz. Forces exerted by both optical traps and the position of the mobile trap center were recorded. From this data, we first determined the system spring constant $\chi^\ast_{sys}(\omega)$ using Eq.~\eqref{eq:SysStiff}  (Fig.~\ref{fig:Fig1}e). The frequency dependent spring constant exhibits the three regimes discussed above (compare Fig.~\ref{fig:Fig1}e and Fig.~\ref{fig:Fig2}c): a low frequency regime dominated by surface tension, an intermediate regime and a high frequency regime dominated by the traps (Fig.~\ref{fig:Fig1}e). \\
We then determined the complex spring constant of uniaxial droplet elongation $\chi^\ast$ using Eq.~\eqref{eq:DroStiff} (Fig.~\ref{fig:Fig3}a).
From this data, we first determined the surface tension $\gamma$ in the low frequency regime using
\begin{equation}
\gamma \approx \chi^\prime(\omega)/(1.25 + 4.36\, \theta_0^2) \quad,
\label{eq:SurfTens}
\end{equation}
valid for small $\theta_0=r_{bead}/R\ll1$ (see Supplement). We then obtain the complex shear modulus $G^\ast$ by accounting for surface tension effects from the droplet spring constant (Fig.~\ref{fig:Fig3}b and Supplement). For modulus $R|G^\ast(\omega)|\gg\gamma $ and small $\theta_0$ we use 
\begin{equation}
G^\ast(\omega)   \approx \frac{(\chi^\ast(\omega)- (1.75 + 6.31\, \theta_0^2)\gamma)}{R (0.58+3.42\, \theta_0^2)  } \quad
\label{eq:ElasMod}
\end{equation}
(see Supplement). 	
Fig.~\ref{fig:Fig3}c-f show storage and loss moduli as a function of frequency averaged over several experiments for four different salt concentrations. Loss moduli $G^{\prime\prime}$ increased approximately linearly for increasing frequency (Fig.~\ref{fig:Fig3}b-f). Corresponding viscosity values range from $\eta=0.1$ to $1\,{\rm Pa}\,$s. Storage moduli at a reference frequency of $10~$Hz ranged between values of $0.1$ and $15$~Pa. Both, viscosities and storage moduli decreased with increasing salt concentration (Fig.~\ref{fig:Fig3}g and i).  The associated loss tangent $G^{\prime\prime}/G^\prime$ exhibited a sharp drop when salt concentration was reduced to 75~mM, suggesting a more solid-like rheology at lower salt concentrations (Fig.~\ref{fig:Fig3}j).
{\color{black}The estimated surface tension $\gamma$ was about $5\,\mu$N/m for $75$~mM salt and decreased steadily for increasing salt concentrations with $\gamma\approx 1~\mu$N/m at $180$~mM salt (Fig.~\ref{fig:Fig3}h and Supplement). }\\
 {\color{black} To test our new method, we measured viscosity and surface tension using two established methods. Using passive  bead-tracking micorheology, we estimated a viscosity $\eta= 0.7\pm 0.1\,$Pa~s at 75 mM salt which is in excellent agreement with our active microrheology results  (Fig.~\ref{fig:Fig3}g and Supplement).  Furthermore, we analysed thermal fluctuations of droplet shape to measure surface tension $\gamma\approx 1.4\,\mu$N/m at 180 mM salt, consistent with our new method.}\\
{\color{black} The surface tension and viscosity of polyelectrolyte condensates have been studied theoretically \cite{spru10, spru10b}. In Fig.~3h, we compare the predicted salt-dependence of surface tension to our data and find good agreement. Furthermore, the salt dependence of a characteristic  time is also in agreement with this theory (see supplemental Fig.~S6).
}

{\it Discussion.} Here, we used a novel technique with optical traps to characterize the rheology of PGL-3 protein droplets. We find that newly formed PGL-3 droplets consist of a viscoelastic material with liquid-like material properties that depend on salt concentration. Viscosities range from $0.1$ to $1\,$Pa~s for decreasing salt concentrations from $180$ to $75$~mM. The storage modulus is smaller than the loss modulus, but approaches the loss modulus from below for frequencies larger than $10$~Hz.
Furthermore, we obtain more precise values of surface tension than previous methods. Values depend on salt concentration and range from 1~$\mu$N/m to 5~$\mu$N/m for decreasing salt concentrations from $180$ to $75$~mM. Such low values of surface tension are consistent with values estimated for colloidal systems with weak interactions  but much smaller than typical air-water or oil-water systems \cite{lekk08, lekk08b}. Notably, our  viscosity and surface tension estimates for PGL-3 droplets {\it in vitro} are consistent with estimates for P-granules {\it in vivo} suggesting that {\it in vitro} PGL-3 is a good model system for the physics and material properties of P-granules {\it in vivo} \cite{bran09}.
Interestingly, our viscosity and surface tension values for PGL-3 are between one to two orders of magnitude smaller than estimates for droplets of the P-granule protein LAF-1 formed {\it in vitro}  \cite{elba15}.\\
Viscosities of PGL-3 droplets exhibit a strong salt dependence qualitatively similar to salt-dependent viscosity estimates reported for phase-separated LAF-1 by passive bead-tracking micro-rheology (Fig.~\ref{fig:Fig3}g) \cite{elba15}.
The decrease in storage moduli, loss moduli and surface tensions for increasing salt concentrations suggests that {\color{black} screened electrostatic interactions play an important role in protein condensates and their material properties. As salt concentration is increased, interactions become weaker because of screening effects.
Interestingly, synthetic polyelectrolytes show a similar dependence of rheology and surface tension on salt concentration as the protein condensates studied here \cite{spru10, spru10b}.  }
\\
In conclusion, our measurements show that protein condensates are complex polymeric liquids with viscoelastic material properties that can be regulated by salt concentration. We obtain our results using a novel optical trap based rheometer suitable for micron-sized probes. This method does not rely on thermodynamic equilibrium. We expect that our work will trigger more studies that lead to a deep understanding of the physics of protein droplets and their phase behavior in biological cells.
\section{Acknowledgments}
We would like to thank Marcus Jahnel and Stephan Grill for useful discussions and the staff of LUMICKS for their technical assistance.
Furthermore, we would like to thank the following Services and Facilities of the MPI-CBG for their support: The Protein Expression Purification and Characterization (PEPC) facility.
 
%

\end{document}


\title{ {\Large   Supplementary Material} \\
{\rm Salt-dependent rheology and surface tension of protein condensates using optical traps}}
\author{Louise Jawerth*}
\affiliation{Max-Planck-Institut f\"ur Zellbiologie und Genetik, Pfotenhauerstra{\ss}e 108, 01307 Dresden, Germany}
\affiliation{Max-Planck-Institut f\"ur Physik komplexer Systeme, N\"othnitzerstra{\ss}e 38, 01187 Dresden, Germany}
\author{Mahdiye Ijavi}
\affiliation{Max-Planck-Institut f\"ur Zellbiologie und Genetik, Pfotenhauerstra{\ss}e 108, 01307 Dresden, Germany}
\author{Martine Ruer}
\affiliation{Max-Planck-Institut f\"ur Zellbiologie und Genetik, Pfotenhauerstra{\ss}e 108, 01307 Dresden, Germany}
\author{Shambaditya Saha}
\affiliation{Max-Planck-Institut f\"ur Zellbiologie und Genetik, Pfotenhauerstra{\ss}e 108, 01307 Dresden, Germany}
\author{Marcus Jahnel}
\affiliation{Max-Planck-Institut f\"ur Zellbiologie und Genetik, Pfotenhauerstra{\ss}e 108, 01307 Dresden, Germany}
\affiliation{Biotechnology Center, Technische Universit\"at Dresden,  Tatzberg 47-49, 01307 Dresden, Germany}
\author{Anthony A Hyman}
\affiliation{Max-Planck-Institut f\"ur Zellbiologie und Genetik, Pfotenhauerstra{\ss}e 108, 01307 Dresden, Germany}
\author{Frank J\"ulicher}
\affiliation{Max-Planck-Institut f\"ur Physik komplexer Systeme, N\"othnitzerstra{\ss}e 38, 01187 Dresden, Germany}
\author{E. Fischer-Friedrich}
\affiliation{Biotechnology Center, Technische Universit\"at Dresden,  Tatzberg 47-49, 01307 Dresden, Germany}
\affiliation{Max-Planck-Institut f\"ur Zellbiologie und Genetik, Pfotenhauerstra{\ss}e 108, 01307 Dresden, Germany}
\affiliation{Max-Planck-Institut f\"ur Physik komplexer Systeme, N\"othnitzerstra{\ss}e 38, 01187 Dresden, Germany}


\maketitle

\section*{{ \large Materials and Methods}}
{\bf Protein preparation.}
We use the protein purification protocol presented in Saha {\it et al.} \cite{saha16}. Untagged PGL-
3 was further purified with HiLoad 16/60 Superdex 200 size-exclusion
chromatography column (GE Healthcare) in size exclusion buffer. Untagged PGL-3
was stored in the same conditions as tagged PGL-3 (storage buffer: 25�mM HEPES pH 7.5, 0.3�M KCl, 1�mM DTT). We added a small fraction of polystrene beads (Thermofisher, FluoSpheres$^{\rm TM}$ F8887) of  a diameter of $1\,\mu$m to a buffer which lacked KCl. The PGL-3 droplet assembly was initiated through dilution of PGL3 in storage buffer using the bead solution to reach the desired, final salt concentration.  This resulted in a buffer solution with many $\sim 10\, \mu$m droplets with associated immersed beads (final buffer: $25$mM HEPES, $1$mM DTT, specified KCL, pH $7.5$). Shortly after assembly, $1-2\,\mu$l of the droplet solution was pipetted between two glass coverslips separated by a $100\,\mu$m spacer. The bottom glass coverslip was PEG-passivated to minimize droplet adhesion. Elsewhere on the same sample chip, we also pipette $1-2\,\mu$l of beads in a buffer which does not contain protein. We use beads in this region to thermally calibrate trap stiffnesses.\\
For the confocal image in Fig.~1b, main text {\color{black} and Movie 1,} we follow the procedure above with the exception that we use protein before the GFP-tag cleavage step. The image is obtained using the built-in confocal of the optical trap and a $488\,$nm laser illumination.\\

{\bf Measurement setup.}
The sample chamber was loaded into the dual-trap optical trap (C-Trap$^{\rm TM}$, LUMICKS) and all measurements were made within two hours of droplet formation. Optical trap stiffness was determined by fitting a Lorentzian function to the power spectral density of the fluctuations of a trapped bead \cite{jone15} using the built-in thermal calibration routine of the optical trap; for this calibration, beads without associated protein droplets were used. Trap stiffnesses were in the range of $140-170\, {\rm pN}/\mu$m for salt concentrations 75-150 mM. We reduced laser powers of the optical trap at  salt concentration 180 mM to avoid laser induced artifacts that we attribute to heating. This resulted in trap stiffnesses of $60-80\,{\rm pN}/\mu$m at 180 mM salt concentration.\\
To measure the rheological properties of a droplet, we trapped a single bead in each of the two traps; both beads were immersed in a large droplet. We moved the trap centers apart until the droplet was slightly stretched. We applied an oscillatory strain at several predefined frequencies to protein droplets while measuring the corresponding force.\\
{\color{black}To illustrate droplet manipulation in the optical trap, a time- lapse of a GFP-tagged droplet being pulled apart with $2 \mu m$ diameter beads is shown in Movie 1.\\ }

{\color{black}
{\bf Trap calibration.}
During the experiments, we calibrated the optical trap with a naked bead inside aqueous buffer solution obtaining trap stiffness $k$ and sensitivity $1/\beta$. We find that the conversion factor  $k\beta$ between force and voltage is largely independent (within 10\%) of bead size and has the same value if a bead is replaced by droplets of different size. This is consistent with previous reports showing that this conversion factor is independent of the specific conditions of the experiment, but is a feature of the instrument \cite{farr12}. 
\\
The trap stiffness $k$  is dependent  on bead size and refractive index. To test whether the presence of the droplet influences the value of $k$ during our measurements, we measure the system spring constant $\chi^\ast_{sys}$ using the conversion factor $k\beta$ (Fig.~1a, main text).
 For the case of the droplet spring constant $\chi^\ast$ being significantly larger in magnitude than the stiffnesses of the optical traps, the droplet provides a quasi-rigid linker between the two beads in the optical traps. Since the displacement of the trap center is imposed, we can, in this case, determine the actual value of the spring constant of the two traps in series $k_1k_2/(k_1+k_2)$ independently as the value of $\chi^\ast_{sys}$.  This situation is expected to be realized for the case of small $k_1, k_2$ and for large values of $\chi^\ast_{sys}$ which we find in the high frequency regime (Fig. 3a, main text). 
Indeed, we observed that the maximal value of $\chi^\ast_{sys}$ converges to $k_{1}k_{2}/(k_{1}+k_{2})$ for spring constants $k_{1},k_{2}\lesssim 60\,{\rm pN}/\mu$m, where $k_1$ and $k_2$ are the spring constants determined from the calibration with naked beads (see Fig.~\ref{fig:HighFreqLowStiff}). If the presence of the droplet material would have altered the trap stiffnesses, these values would not have agreed. In summary, this shows that both the conversion factor and the trap stiffness are largely unchanged by the presence of the droplet within the measurement error. \\
\begin{figure*}[h]
\centering
\includegraphics[width=7cm]{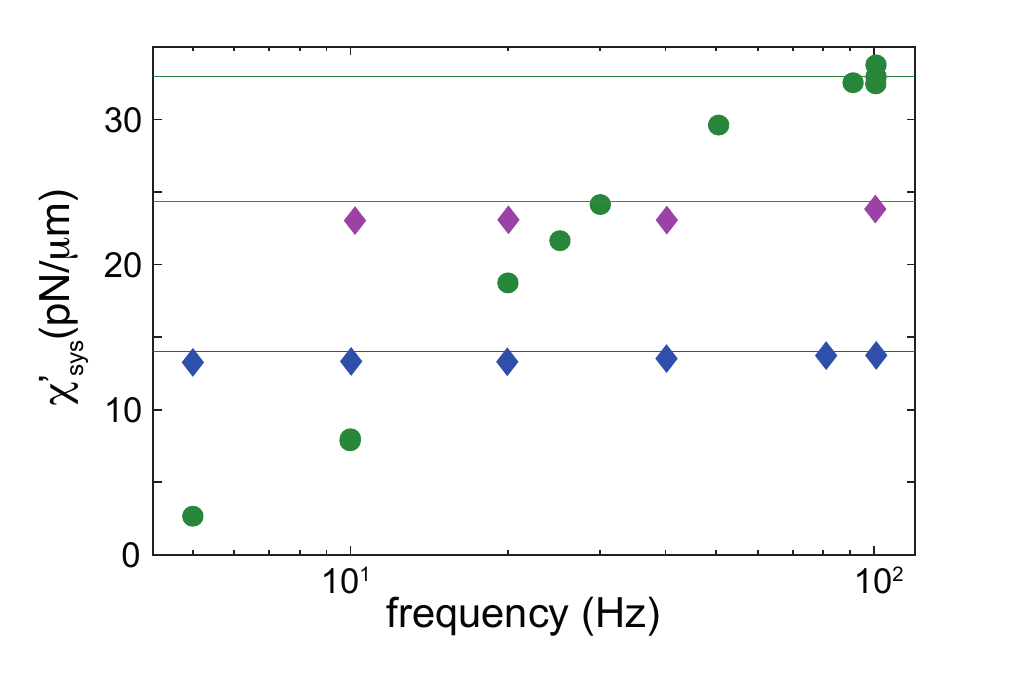}
\caption{ \label{fig:HighFreqLowStiff}
At high frequencies, the system spring constant $\chi^\ast_{sys}$ (rhombi: 75 mM KCL,  circles: 200 mM KCl) approaches the predicted limit value of the combined traps in series $k_1k_2/(k_1+k_2)$  (solid lines, green: $k_1=k_2=66\,{\rm pN}/\mu$m, pink: $k_1=k_2=49\,{\rm pN}/\mu$m, blue: $k_1=k_2=28\,{\rm pN}/\mu$m) as the droplet spring constant becomes considerably larger than trap stiffnesses. Spring constants were determined from the calibration of naked beads.}
\end{figure*}
}

{\bf Data analysis.}
During measurements, the  position of the movable trap center $x_{1}$ and the forces exerted by the two traps  $F_1$ and $F_2$ were recorded. It is important to note that $x_1=C+ x_{sys}$ (Fig. 1c, main text), where $C$ is a constant. Therefore, the time-periodic change in the system's length $x_{sys}$ equals the time-periodic change in $x_1$ during measurements. 
The inter-bead distance $d_b$ between the two trapped beads was obtained through a LUMICKS C-trap built-in image analysis algorithm. 
From the time-average of the inter-bead distance $<d_b>$, we determined the average protein droplet radius as  $R=<d_b>/2+ r_{bead}$, where $r_{bead}$ was the bead radius.
The corresponding droplet radius $R$ was confirmed by brightfield images of droplets.\\
We first determined the one-dimensional rheological response of the two traps in series with the droplet using $\sigma=(F_1-F_2)/2$ as stress variable. The corresponding time-periodic stress signal and the time-periodic trap distance $x_{sys}$ were fit by sinusoidal curves to determine the amplitudes $A_\sigma$ and $A_{x_{sys}}$ and phase $\varphi_\sigma$ and $\varphi_{x_{sys}}$ of either signal. The spring constant of the system $\chi_{sys}^\ast$  was then calculated as $A_\sigma/A_{x_{sys}} \exp(i(\varphi_\sigma-\varphi_{x_{sys}}))$ (Fig.~1a, main text). 
We then calculated the droplet spring constant $\chi^\ast$ from Eq.~(5), main text. The hydrodynamic drag coefficient of the droplet was estimated as the Stokes drag $\xi=6\pi R \eta_b $, where  $\eta_b=10^{-3}\,$Pa~s is the viscosity of the buffer. 
Using the droplet spring constant $\chi^\ast$, we identified (by eye) a low frequency regime, where the storage modulus dominates and is constant. In this regime, our theoretical considerations  indicate that surface tension effects dominate (Fig.~2a,b, main text). For our measurements, this regime was identified as $f\le 0.2\,$Hz (Fig.~3a, main text). We averaged the measured storage moduli in this frequency regime and determined surface tension by use of Eq.~\eqref{eq:SurfTens}. There, we  set $\theta_0=r_{bead}/R$ where $r_{bead}$ is the radius of the trapped polystyrene beads and $R$ is the radius of the droplet. 
Furthermore, we identified a high-frequency regime where the real part of the droplet spring constant $\chi^\ast$ significantly exceeds the surface-tension-plateau  (Fig.~3a, main text). For our measurements, this regime was identified as $f\ge 4$~Hz.
In this high-frequency regime, we used the expansion Eqn.~\eqref{eq:ElasMod} to calculate $G^\ast$ (see also Eq.~10, main text). In addition, we determined the loss modulus of the droplet for the frequency range $f< 4\,$Hz by numerically solving Eq.~\eqref{eq:LossModOnly} as in this regime $G^\prime\ll G^{\prime\prime}$.\\

{\color{black}
{\bf Passive bead-tracking microrheology.}
We performed passive bead-tracking microrheology to measure viscosity independently by an established method.
We quantified the thermal motion of immersed microspheres ($1\,\mu$m diameter) in large droplets of protein condensate at an exemplary salt concentration of 75 mM (Fig.~\ref{fig:BeadTracking}a). To this end, we generated large protein droplets as described in the subsequent paragraph.
By analyzing the thermal motion of immersed beads over time, we obtain a mean-square-displacement $<r^2(\tau)>$  (Fig.~\ref{fig:BeadTracking}b). We find that the time-dependence of $<r^2(\tau)>$ is captured well by a linear fit with a slope of $4D\approx 0.0025\,\mu$m/s, where $D$ is the diffusion coefficient. From the Stokes-Einstein relation, we find $ D= k_B T/(6\pi \eta r)$, where $D$ is the diffusion constant and $r$ is the radius of the bead. The corresponding viscosity estimate is $\eta\approx 0.7\,$Pa~s of the protein condensate at 75 mM salt concentration.\\
{\it Experimental details:} 
We loaded $25 \mu l$ of PGL-3 in a high salt ($300$~mM KCl) buffer into a custom-made chamber that incorporates a dialysis membrane; this chamber allows buffer exchange while simultaneously imaging the sample through a glass bottom. Before addition to the chamber, we also added a small fraction of $1 \mu m$ diameter red-fluorescent beads (Thermofisher, FluoSpheres$^{\rm TM}$ F8887) to the protein solution. Both the beads and glass coverslip were coated with poly-ethylene glycol (PEG) to minimize adhesion between these surfaces and the protein. Droplet assembly was initiated by changing buffer conditions such that there is a final salt concentration of $75$~mM KCl. We wait $\sim 30$ min to allow most coalescence events to cease. Subsequently, we acquire image stacks representing the three-dimensional volume surrounding a large ($\sim 30\,\mu$m radius) droplet  with a voxel size of $0.136$ x $0.136$ x $0.4\,\mu$m at a rate of $4$ seconds per stack for over $2$ hours. To analyze the motion of the beads, we created a maximum projection in the Z-direction of the droplet mid-section for each image stack acquired; for an imaging volume with a small number of bright points, this results in an image that represents the projection of features onto the $x\mbox{-}y$ plane, effectively removing the $z$-component of their motion. We repeat this procedure at every time point to create a movie of the bead motion in $x\mbox{-}y$.
We use the TrackMate algorithm (through Fiji plugin) to locate particle positions over time from the aligned image stack  using sub-pixel localization resulting in a list of $x,y$ positions over waiting time. We only analyze bead tracks that are longer than $1000$ frames ($4000$~s). Moreover, we identify all beads that are close to the surface of the droplet and do not consider these for bead diffusion analysis.\\
\begin{figure*}[h]
\centering
\includegraphics[width=12cm]{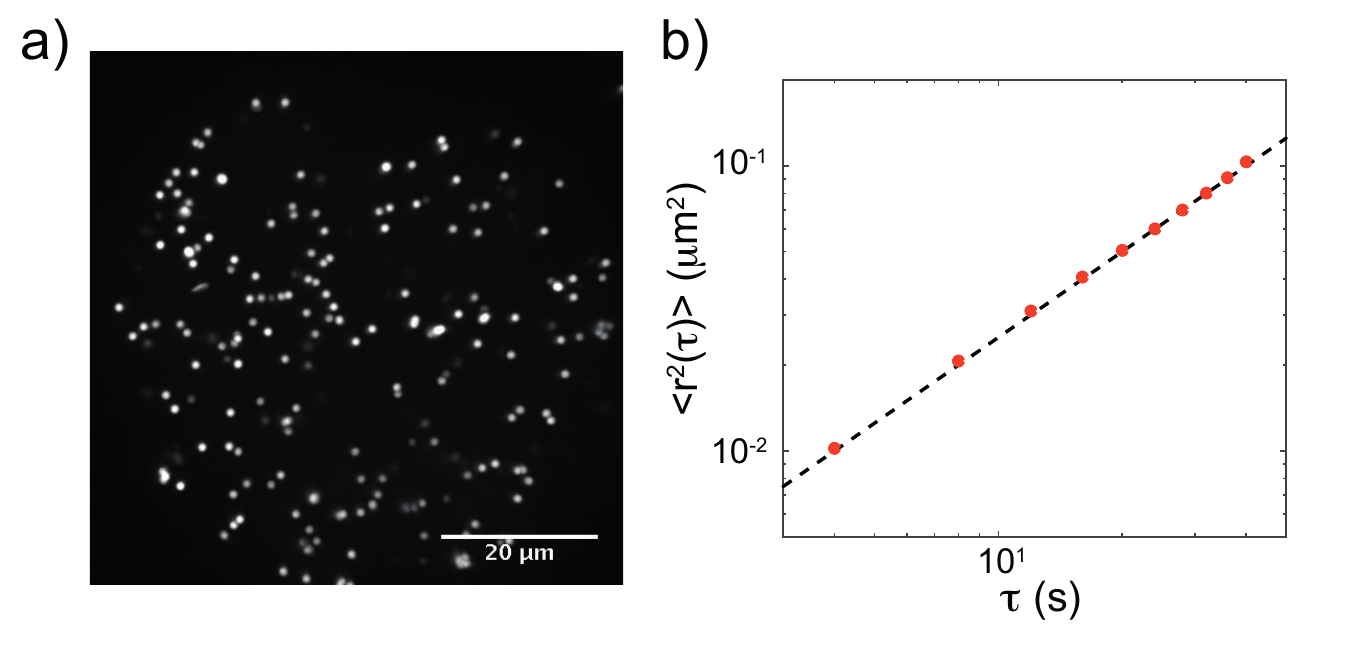}
\caption{ \label{fig:BeadTracking}
\color {black} Thermal motion of immersed microspheres (diameter: $1\,\mu$m). a) Maximum intensity projection of a confocal z-stack of the mid-section of a protein droplet (not directly visible) with immersed fluorescent beads at salt concentration of $75\,$mM. b) Mean squared displacement of microspheres inside the protein droplet (red data points). The dashed black line shows a linear fit $4 D \tau$ with a slope of $4D\approx 0.0025\,\mu{\rm m}^2$/s.
}
\end{figure*}

{\bf Analysis of droplet shape fluctuations.}
We also performed an analysis of droplet shape fluctuations to measure droplet surface tension independently by an established method.
For this purpose, we quantified the Fourier modes of capillary waves of a protein droplet at a salt concentration of $180\,$mM. At high salt concentration, noticeable droplet shape fluctuations are present (see Movies 1 and 2). From a time-lapse brightfield movie of a droplet, we analysed droplet shape fluctuations normal to the surface boundary. The movie of a total time of $5.91\,$s was recorded with a 60x water immersion objective (NA 1.2) at a time interval of 0.01 s and a pixel size of 90 nm.
We then performed image processing on this movie using Fiji \cite{schi12} (see Fig.~\ref{fig:ImageProcessing}): First, we inverted the image and then binarized it by using the `Auto local Threshold' plugin (Method: Bernsen, radius: 10, parameter 1: 20, parameter 2: 0). To the obtained binary image, we applied the `Analyze Particles' plugin detecting the outline of `particles' with a size larger than $50\,$pixel$^2$. The obtained (inner) top and bottom outline of the droplet shape was then used to detect Fourier modes of boundary fluctuations in direction normal to the surface. \\
\begin{figure*}[h]
\centering
\includegraphics[width=14cm]{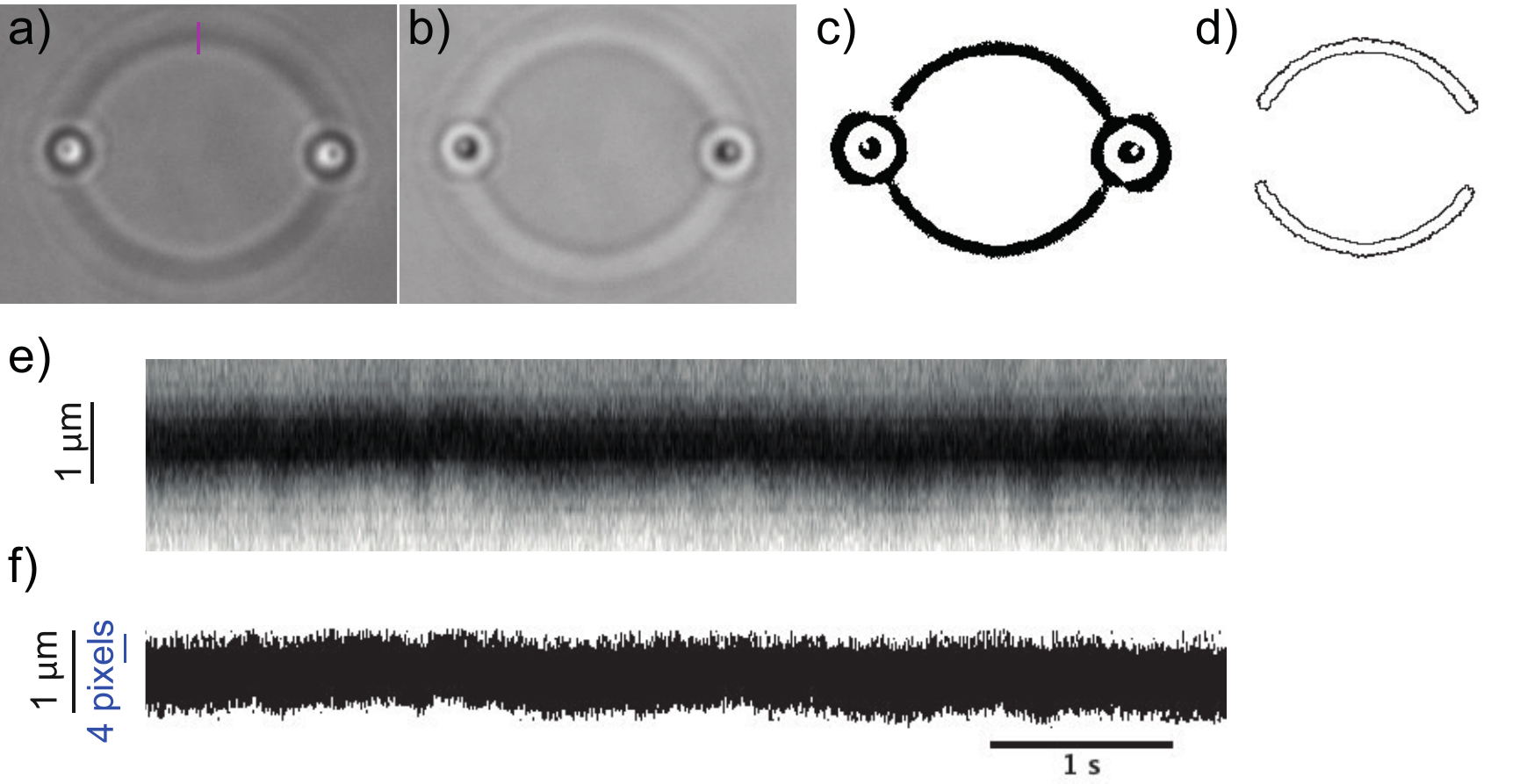}
\caption{ \label{fig:ImageProcessing}
\color{black} Illustration of image processing steps of droplet micrographs (a-d) and droplet boundary fluctuations (e-f). a) Original image. b) Inverted image. c) Binarized  image after local thresholding. 
d) Outlines obtained from particle analysis.
e) Illustration of boundary fluctuations over time from a line scan along the pink line depicted in panel a.
f) Thresholded version of boundary fluctuations from d.
}
\end{figure*}
Shape fluctuations of an unperturbed spherical droplet of radius $R$ may be described as an expansion in spherical harmonics \cite{ljun84}
\begin{align}
r(\theta,\varphi, t)=R\left(1+\sum_{l=2}^{l=l_{max}} \left\{ a_{l,0}(t)Y_{l,0}(\theta, \varphi)
+ \sum_{m=1}^{l}  \left[ \alpha_{l,m}(t)\left((-1)^m Y_{l,m}+Y_{l,-m}\right) +  \frac{\beta_{l,m}(t)}{i}\left((-1)^m Y_{l,m}-Y_{l,-m}\right) \right]
\right\}\right) \label{eq:RadFluc},
\end{align}
where $Y_{l,m}(\theta, \varphi)$ is the spherical harmonics function of degree $l$ and order $m$.
Surface fluctuations give rise to fluctuations in the surface energy  \cite{ljun84,betz12}
\begin{align*}
\delta\Phi=\frac{1}{2}\gamma R^2 [l(l+1)-2]\left(a_{l,0}^2 +\sum_{m=1}^{l}  (\alpha_{l,m}^2+\beta_{l,m}^2)\right).
\end{align*}
According to the equipartition theorem, we have \cite{ljun84}
$$
<\alpha_{lm}^2>=<\beta_{lm}^2>=\frac{k_B T}{(\gamma R^2 [l(l+1)-2])}.
$$

For the situation of a droplet held laterally by trapped, adherent beads, we assume that modes which have a non-vanishing amplitude at the position of the beads ($\theta=0, \pi$) vanish. This assumption is corroborated by our analysis of surface fluctuations from the recorded time lapse movie. Therefore, we set $a_{l,0}=0$. 
According to Eq. \eqref{eq:RadFluc}, we find for the radius fluctuations along the cross-section of  $\varphi=0$
\begin{align*}
\delta r(\theta,0, t)=R\sum_{l=2}^{l=l_{max}}  \sum_{m=1}^{l} (-1)^m 2 \alpha_{l,m}Y_{l,m}(\theta,0)  
\end{align*}
Let the Fourier decompisition of $\delta r(\theta, 0,t)$ be
 $$\delta r(\theta, 0,t)=\sum_{n=1}^\infty \left(a_n(t)  \cos(2 n \theta)+ b_n(t) \sin(2 n\theta) \right)$$ and let  $c_n=\sqrt{a_n^2+b_n^2}$.
 Then, the time-average  $<c_{n}^2>$  is
\begin{align*}
<c_{n}^2>&=R^2\sum_{l=2}^{l=l_{max}}  \sum_{m=1}^{l}  4 <\alpha_{l,m}^2>  \{ F_{n,l,m,1}^2+F_{n,l,m,2}^2\} \\
&= \sum_{l=2}^{l=l_{max}} \frac{ 4 k_B T}{(\gamma [l(l+1)-2])} \sum_{m=1}^{l}   \{ F_{n,l,m,1}^2+F_{n,l,m,2}^2\},
\end{align*}
with $F_{n,m,l,1}=\frac{2}{\pi} \int_0^\pi Y_{l,m}(\theta,0) \cos(2 n\theta) {\rm d}\theta$ and $F_{n,m,l,2}=\frac{2}{\pi} \int_0^\pi Y_{l,m}(\theta,0) \sin(2 n\theta) {\rm d}\theta$ the Fourier coefficients of the function $Y_{l,m}(\theta,0)$.
For our analysis, we approximated $<c_{n}^2>$ as
$$
<c_{n}^2>\approx \sum_{l=2}^{l=15} \frac{ 4 k_B T}{(\gamma [l(l+1)-2])} \sum_{m=1}^{l}   \{ F_{n,l,m,1}^2+F_{n,l,m,2}^2\}
$$
for Fourier coefficients up to $n=5$.
The obtained values of $<c_{n}^2>$ and the associated  estimates of $\gamma$ as 
$$
\gamma\approx \frac{1}{<c_n^2>} \sum_{l=2}^{l=15} \frac{ 4 k_B T}{([l(l+1)-2])} \sum_{m=1}^{l}   \{ F_{n,l,m,1}^2+F_{n,l,m,2}^2\}
$$
are depicted in Fig.~\ref{fig:ShapeFluc}a and b, respectively. Our estimates of $<c_n^2>$ are well above the noise floor, which we estimate to be $ 300\,{\rm nm}^2$  from the apparent boundary fluctuations of the imaged bead in our time lapse movie. The obtained surface tension estimates of $\approx 1.4\,\mu$N/m are indeed largely constant and agree to a good extent with the corresponding tweezer measurements  at salt concentration $180\,$mM which gave a surface tension estimate of $\approx 1\,\mu$N/m. 
From the decay of the autocorrelation function of the Fourier coefficients $a_n$ and $b_n$, we estimate a characteristic decay time  $\tau_{relax}\approx 0.13\,$s (Fig.~\ref{fig:ShapeFluc}c). We know that $\tau_{relax}\propto \gamma/(\eta R)$, where $R$ is the droplet radius and $\eta$ its viscosity. This allows to make an order of magnitude estimate of $\eta$ as $\tau_{relax}\gamma/ R$. 
Measuring $R\approx 6.9\,\mu$m, we obtain $\eta\approx 0.026\,$Pa~s, which is indeed on the same order of magnitude as our tweezer measurement result of $0.07\,$Pa~s at salt concentration of 180 mM (see Fig.~3g, main text).
\\
We believe that our estimate of surface tension through capillary wave analysis is a coarse estimate of surface tension as our analysis predicts mean squared fluctuations of the shape that are well  below the limit of optical resolution $(<c_1^2>\approx (40\,{\rm nm})^2)$. 
In this case, Fourier mode analysis of capillary waves most likely underestimates values of $<c_n^2>$ and thus overestimates $\gamma$.
\begin{figure*}[h]
\centering
\includegraphics[width=16cm]{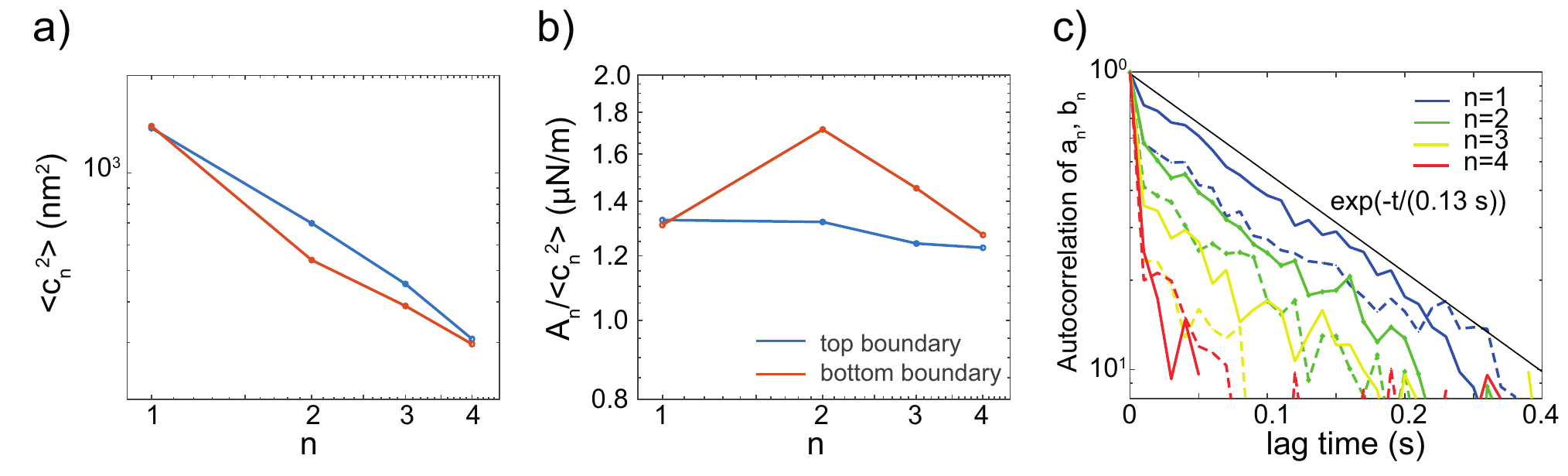}
\caption{ \label{fig:ShapeFluc}
\color{black} Shape fluctuations of a protein droplet at salt concentration $180\,$mM. 
a) Time-averaged squared Fourier coefficients $c_n$ of shape fluctuations normal to the droplet boundary.
b) Surface tension estimate for different Fourier coefficients according to the formula $A_n/<c_n^2>$ where $A_n=\sum_{l=2}^{l=15} { 4 k_B T}/{([l(l+1)-2])} \sum_{m=1}^{l}   \{ F_{n,l,m,1}^2+F_{n,l,m,2}^2\}$.
c) Autocorrelation of the Fourier coefficients $a_n$ (solid line) and $b_n$ (dashed line). We find a characteristic slowest decay time of $\approx 0.13\,$s.
}
\end{figure*}

}

\section*{\large Continuum mechanics theory of optical trap microrheology}
In this section, we estimate the deformation response of a micron-sized spherical viscoelastic droplet immersed in a surrounding liquid of significantly lower viscosity. The droplet  is subject to an oscillatory pressure applied with angular frequency $\omega$ in two dome-shaped regions at its poles (Fig.~2a, main text). We will determine the complex frequency-dependent spring constant of uniaxial droplet deformation $\chi^\ast(\omega)$ in dependence of surface tension $\gamma$ and complex shear modulus $G^\ast(\omega)$ of the droplet material. 
For this purpose, we will perform a perturbation equation to first order in stress, strain and deformation. 
We choose the coordinate system such that the problem is rotationally symmetric with respect to the z-axis (Fig. 2, main text) and there is no dependence of the displacement on the azimuthal angle $\varphi$. 
The radial and axial displacement of the droplet $u_r$ and $u_{\theta}$, the associated internal velocities $v_r$ and $v_\theta$ and  the internal pressure change  $\delta p$ are time-dependent. As we are interested in the calculation of the displacement and the velocities to first order, the governing equations are linear. Thus, time dependence can be factorised and we can write
\begin{eqnarray}
u_{r,\theta}( r, \theta, t)&=&\tilde{ u} (r, \theta)  \exp(i \omega t)+c.c.\\
v_{r,\theta}( r, \theta, t)&= & i \omega \tilde{u}_{r,\theta} (r, \theta) \exp(i \omega t)+c.c.\\
\delta p(r,\theta, t)&=&\delta \tilde p(r,\theta)\exp(i\omega t)+c.c. \quad.
\end{eqnarray}
Here, the tilde denotes the complex amplitude of the Fourier mode at angular frequency $\omega$ and $c.c.$ refers to the complex conjugate.
In the following, we will leave out the tilde above $u$, $v$ and $p$ keeping in mind that the time dependence of these functions is captured by a simple time-periodic modulation. Furthermore, we will omit the  expression ``$+ c.c.$'' for convenience.  \\
The Reynolds number of the hydrodynamic problem under consideration is $Re=\rho v  R/ \eta$ where $R$ is the droplet radius. The fluid velocity may be estimated as $v\approx \omega R$. Assuming a droplet of $R\le 10\,\mu$m, a density similar to water $\rho\approx 1000\,kg/m^3$, frequencies $f\le 100\,$Hz and $\eta >10^{-3}\,Pa\,s$, we obtain as an upper bound estimate for the Reynolds number $Re\lesssim 10^{-2}$ which is $\ll 1$.
Therefore, we will consider the dynamics of the fluid constituting the droplet as over-damped. \\ 
Force balance and incompressibility requires (in Cartesian coordinates) $\partial_{i}\sigma_{ij}=0$ and $\partial_i v_i=0$.
For the viscoelastic stresses inside the incompressible droplet, we make the ansatz (in Cartesian coordinates)
\begin{equation}
\sigma_{ij} = - p\, \delta_{ij}+G^\ast(\omega)  \left[\partial_i  u_j+\partial_j  u_i\right] \quad, 
\end{equation}
where $G^\ast(\omega)$ is the frequency-dependent complex elastic shear modulus of the droplet material. Note, that if the droplet is a Newtonian liquid with $G^{\ast}(\omega)=i\omega\eta$, the force balance equation equals the Stokes equation  in Fourier space as  $\underline{v}=i\omega\underline{u}$. In spherical coordinates, we obtain\cite{ache90}
\begin{align}
\  &\frac{\partial p}{\partial r}  =
 G^\ast(\omega) \left[\frac{1}{r^2} \frac{\partial}{\partial r}\left(r^2 \frac{\partial u_r}{\partial r}\right) +
                   \frac{1}{r^2 \sin(\theta)} \frac{\partial}{\partial \theta}\left(\sin(\theta) \frac{\partial u_r}{\partial \theta}\right) - 2\frac{u_r +
                  \frac{\partial u_{\theta}}{\partial \theta} + u_{\theta} \cot(\theta)}{r^2}
                              \right] \\
\  & \frac{1}{r} \frac{\partial p}{\partial \theta}  =
         G^\ast(\omega) \left[\frac{1}{r^2} \frac{\partial}{\partial r}\left(r^2 \frac{\partial u_{\theta}}{\partial r}\right) +
                         \frac{1}{r^2 \sin(\theta)} \frac{\partial}{\partial \theta}\left(\sin(\theta) \frac{\partial u_{\theta}}{\partial \theta}\right) +
                       \frac{2}{r^2} \frac{\partial u_r}{\partial \theta} - \frac{u_{\theta} }{r^2 \sin(\theta)^2}
                 \right].
\end{align}
\normalsize
As the viscosity of the droplet environment is assumed to be significantly lower than the droplets viscosity and velocities have to match at the phase boundary, viscous stresses outside the droplet are negligible and we set them to zero. 
Force balance at the droplet boundary imposes 
\begin{align}
-p_{ext}- \sigma_{rr}(R,\theta) & = 2 \gamma H \label{eq:bound1},\\
\sigma_{r,\theta}(R,\theta) &=0 \label{eq:bound2}
\end{align}
where $p_{ext}$ is the local normal force per area applied on the droplet externally through the oscillatory forcing at the poles and an external hydrostatic pressure. Furthermore, $\gamma$ is the droplet's surface tension that gives rise to a pressure jump across the droplet  boundary. The mean curvature of the droplet is denoted by $H$. If the droplet is unperturbed, it acquires a spherical shape with mean curvature $H=1/R$. In this case, Eq.~\eqref{eq:bound1} states Laplace's law  in the form 
$-p_{ext}+p  = 2 \gamma/R$, where $p_{ext}$ is a constant external hydrostatic pressure.
If the droplet is slightly deformed from the spherical shape in terms of a displacement $(r,\theta)\rightarrow (r+u_r, \theta+u_\theta)$ through oscillatory forcing at the poles, one may express the change in the mean curvature of the droplet surface to first order in the displacement by $- (2u_r+\partial_\theta u_r\cot\theta+\partial^2_\theta u_r)/(2R^2)$ \quad \cite{miet15}.\\
Substituting $\sigma_{rr} =- p+2G^\ast(\omega) \partial_r u_r $ and 
$\sigma_{r,\theta}(r,\theta)= G^\ast(\omega)\left(\frac{1}{r}\frac{\partial u_r}{\partial \theta}+\frac{\partial u_\theta}{\partial r}-\frac{u_\theta}{r} \right)$ \quad \cite{land87}
  into Eqn.~\eqref{eq:bound1},\eqref{eq:bound2}, we obtain
\begin{align}
-\delta p_{ext} &=\left(-\gamma (2u_r+\partial_\theta u_r\cot\theta+\partial^2_\theta u_r)/R^2 - \delta p+2G^\ast(\omega) \partial_r u_r \right) \big{|}_{r=R} \label{eq:bound1B}\\
0 &=G^\ast(\omega)\left(\frac{1}{r}\frac{\partial u_r}{\partial \theta}+\frac{\partial u_\theta}{\partial r}-\frac{u_\theta}{r} \right)\Big{|}_{r=R}
, \label{eq:bound2B}
\end{align}
where $\delta p$ and $\delta p_{ext}$ are the change of the internal and external pressure change, respectively.
To solve Eqn.~(\ref{eq:bound1}, \ref{eq:bound2}), we use a stream function approach
\begin{align*}
i\omega u_r&=-\frac{1}{r^2 \sin(\theta)}\frac{\partial \psi}{\partial \theta} ,\quad
i\omega u_\theta=\frac{1}{r \sin(\theta)}\frac{\partial \psi}{\partial r}.
\end{align*}
A solution to Eqn.~(\ref{eq:bound1}, \ref{eq:bound2}) with the correct symmetries of the problem  is given by the ansatz \cite{habe58}
\begin{align*}
\psi(r,\theta)&=\sum_{n=3, n\, odd}^\infty \frac{P_{n-2}(\cos\theta)-P_{n}(\cos\theta)}{2n-1} \left(A_n \tfrac{r^n}{R^{n-2}} +C_n \tfrac{r^{n+2}}{R^{n}}\right)
\end{align*}
giving rise to the solutions 
\begin{align}
i\omega u_r=v_r&= -\sum_{n=3, n\, odd}^\infty P_{n-1}(\cos\theta) \left(A_n \tfrac{r^{n-2}}{R^{n-2}}+ C_n \tfrac{r^{n}}{R^n}\right)   \label{eq:urExp}\\
i\omega u_\theta=v_\theta&= \sum_{n=3, n\, odd}^\infty \frac{P_{n-2}(\cos\theta)-P_{n}(\cos\theta)}{(2n-1)\sin\theta}\left(n A_n \tfrac{r^{n-2}}{R^{n-2}}+(n+2)C_n \tfrac{r^{n}}{R^{n}}\right) \notag\\
\delta p &= -\frac{ G^\ast(\omega)}{i\omega}  \sum_{n=3, n\, odd}^\infty P_{n-1}(\cos\theta) \frac{2(2n+1)}{(n-1)} C_n \tfrac{r^{n-1}}{R^{n}}   \notag
\end{align}
 where $P_n$ is the Legendre polynomial of $n$-th degree.
This solution of the flow field inside the sphere needs to be matched with the boundary conditions given by Eqn.~\eqref{eq:bound1B} and \eqref{eq:bound2B} for the applied change of external pressure $\delta p_{ext}(\theta)$ by an adequate match of the coefficients $A_n$ and $C_n$. To achieve this, an expansion of the external pressure change in terms of Legendre polynomials is calculated
 $$\delta p_{ext}=\sum_{n=3, n\, odd}^\infty X^{n-1} P_{n-1}(\cos\theta)  $$ where 
 $$ X^{n}=\frac{2n+1}{2} \int_0^\pi \delta p_{ext}(\theta) P_{n-1}(\cos\theta) \sin(\theta) {\rm d}\theta\quad.$$
For the problem under consideration, the  external pressure is given by an even pressure applied at two symmetric caps at the poles of the sphere. Thus,  
$$\delta p_{ext}(\theta)=P_0\big\{\Theta(\theta-\pi +\theta_0)+\Theta(\theta_0-\theta)\big\}.$$
We have solved the system of equations for $A_n$ and $C_n$ up to index $n=11$, truncating the expansion after this. 
As solutions, we obtain
\begin{align}
A_3 &= \frac{(20 P_0 R^2 \omega \cos(\theta_0) \sin(\theta_0)^2)}{i(-20  \gamma - 19  R  G^{\ast}) } \label{eq:ExpCoeff}\\
C_3 &= \frac{ (15 P_0 R^2 \omega \cos(\theta_0) \sin(\theta_0)^2)}{i( 40 \gamma +  38 R  G^{\ast})} \notag\\
A_5 &= \frac{ ( 3 P_0 R^2 \omega \cos(\theta_0) (1 + 7 \cos(2 \theta_0)) \sin(\theta_0)^2)}{i(-36  \gamma - 17 R  G^{\ast}), }\notag\\
C_5 &= \frac{ -((15 P_0 R^2 \omega \cos(\theta_0) (1 + 7 \cos(2 \theta_0)) \sin(\theta_0)^2)}{(8i (-36  \gamma - 17 R  G^{\ast}))) }\notag\\
A_7 &= \frac{ (39 P_0 R^2 \omega \cos(\theta_0) (19 + 12 \cos(2 \theta_0) + 33 \cos(4 \theta_0)) \sin(\theta_0)^2)}{(40 i (-104  \gamma - 33 R  G^{\ast}))}\notag\\
C_7 &= \frac{ -((91 P_0 R^2 \omega \cos(\theta_0) (19 + 12 \cos(2 \theta_0) + 33 \cos(4 \theta_0)) \sin(\theta_0)^2)}{( 128 i (-104  \gamma - 33 R  G^{\ast})))}\notag\\
A_9 &= \frac{ (85 P_0 R^2 \omega \cos(\theta_0) (178 + 869 \cos(2 \theta_0) + 286 \cos(4 \theta_0) + 715 \cos(6 \theta_0)) \sin(\theta_0)^2)}{(448 i (-680  \gamma 
		- 163 R  G^{\ast}))}\notag\\
C_9 &= \frac{ -((153 P_0 R^2 \omega \cos(\theta_0) (178 +  869 \cos(2 \theta_0) + 286 \cos(4 \theta_0) +  
  		  715 \cos(6 \theta_0)) \sin(\theta_0)^2)}{(1024 i (-680 \gamma - 163 R G^{\ast})))}\notag\\
A_{11} &= \frac{ (175 P_0 R^2 \omega \cos(\theta_0) (2773 + 2392 \cos(2 \theta_0) +  5252 \cos(4 \theta_0) + 1768 \cos(6 \theta_0) + 
        4199 \cos(8 \theta_0)) \sin(\theta_0)^2)}{(36864 i (-140  \gamma - 27 R G^{\ast}))}\notag\\
C_{11} &= \frac{ -((385 P_0 R^2 \omega \cos(\theta_0) (2773 + 2392 \cos(2 \theta_0) + 5252 \cos(4 \theta_0) + 1768 \cos(6 \theta_0) + 
          4199 \cos(8 \theta_0)) \sin(\theta_0)^2)}{(98304 i (-140  \gamma - 27 R  G^{\ast})))}  \notag
 \end{align}
 \normalsize
  To calculate the effective one-dimensional rheology of the droplet during oscillatory deformations along the z-axis, we determine the amplitude  of the force  applied to the left and right pole $P_0 2\pi R^2(1-\cos(\theta_0))$  and the (complex) amplitude of the radial displacement $u_r(R,\theta=0)$. 
 The one-dimensional complex elastic modulus of uniaxial droplet deformation is then 
\begin{equation}
\chi^\ast(\omega)=P_0 2\pi R^2  (1-\cos(\theta_0))/(2 u_r(R,0)) \label{eq:G1D_from_ur} \quad,
\end{equation}
where we have used that $x_{d}=2(R+ u_r(R,0))$ (Fig.~1c, main text). 
We obtain an expression for $\chi^\ast(\omega)$ in terms of the complex elastic modulus $G^\ast(\omega)$ of the droplet material  and its surface tension $\gamma$ using Eqn.~\eqref{eq:urExp}, \eqref{eq:ExpCoeff} and \eqref{eq:G1D_from_ur}. While the full expression for $\chi^\ast (\omega)$ in dependence of $G^\ast(\omega)$ and $\gamma$ is too voluminous to be printed here, we will give in the following explicit expressions for several limit cases.\\

{\bf 1) Small complex elastic modulus}\\
 If surface tension $\gamma$ is dominant in the effective one-dimensional storage modulus, one obtains the relation 
\begin{equation}
\chi^\ast(\omega)\approx \frac{ \gamma \csc(\theta_0)^2 (-9.98 + 4.98 \sec(\theta_0))}{2.32 +  \cos(2 \theta_0) + 0.42 \cos(4 \theta_0) + 0.17 \cos(6 \theta_0) + 
   0.08 \cos(8 \theta_0)}
\end{equation}
which simplifies for small $\theta_0 $ to 
\begin{equation}
\chi^\ast(\omega)\approx  \gamma (1.25 + 4.36 \theta_0^2),
\label{eq:SurfTens}
\end{equation}
including contributions of $\theta_0$ up to second order.\\

{\bf 2) Small surface tension}\\
 If the influence of droplet surface tension $\gamma$ is entirely negligible, our calculation yields
\begin{equation}
\chi^\ast (\omega)=  \frac{25.36  \sec(  \theta_0/2)^2 \sec(\theta_0) \, R G^\ast(\omega)}{16.23 + 
   13.68 \cos(2 \theta_0) + 8.01 \cos(4 \theta_0) + 
   3.73 \cos(6 \theta_0) + 2.09 \cos(8 \theta_0)}. \label{eq:NeglSurfTens}
\end{equation}
For small values of $\theta_0$ and $\gamma$, we may perform a Taylor expansion of $\chi^\ast(\omega)/(R|G^\ast|)$ including contributions to first and second order, respectively, in $\gamma/(R|G^\ast|)$  and  $\theta$. This expansion yields the approximate relation
\begin{equation}
\chi^\ast(\omega)\approx (0.58+3.42\, \theta_0^2) R G^\ast(\omega)  +  (1.75 + 6.31\, \theta_0^2)\gamma  
\label{eq:ElasMod}
\end{equation}

{\bf 3) Dominantly fluid-like droplet rheology}\\
 For the case $G^\prime \ll G^{\prime\prime}$ and small $\theta_0$, we obtain the relation
\begin{align}
\chi^{\prime\prime}\approx \Big[&{R \gamma^{16} \left(8.18\cdot 10^{2}+4.16\cdot 10^{3} \theta_0^2\right) {G^{\prime\prime}} }\hspace*{-0.15cm}+\!
{R^3 \gamma^{14} \left(1.09\cdot 10^{3}+5.81\cdot10^{3} \theta_0^2\right) {G^{\prime\prime}}^3}\hspace*{-0.15cm}+ \!
{R^5 \gamma^{12} \left(5.59\cdot 10^{2}+3.09\cdot 10^{3} \theta_0^2\right) {G^{\prime\prime}}^5}
+ \notag\\
&{R^7 \gamma^{10} \left(1.46\cdot 10^{2}+8.22\cdot 10^{2} \theta_0^2\right) {G^{\prime\prime}}^7} +
{R^9 \gamma^8.    \left(21.3                   +1.22\cdot 10^{2} \theta_0^2\right) {G^{\prime\prime}}^9}+
{R^{11} \gamma^6 \left(1.8                     +10.5 \theta_0^2\right) {G^{\prime\prime}}^{11}}+\notag\\
&\hspace{-1cm}{R^{13} \gamma^4 \left(8.78\cdot 10^{-2}\hspace*{-0.15cm}+5.13\cdot 10^{-1} \theta_0^2\right) {G^{\prime\prime}}^{13}} \hspace*{-0.15cm} +\!
{R^{15} \gamma^2 \left(2.26\cdot 10^{-3}\hspace*{-0.15cm} +\! 1.33\cdot 10^{-2}\theta_0^2\right) {G^{\prime\prime}}^{15}} \hspace*{-0.15cm} + \! R^{17} \left(2.39\cdot 10^{-5}\hspace*{-0.15cm} + \! 1.41\cdot 10^{-4} \theta_0^2\right) {G^{\prime\prime}}^{17} \Big]\notag \\
&\hspace{2cm}/\Big[31.83 \gamma^8+22.54 R^2 \gamma^6 {G^{\prime\prime}}^2  +
 4.35 R^4 \gamma^4 {G^{\prime\prime}}^4 + 0.3 R^6 \gamma^2 {G^{\prime\prime}}^6+6.41\cdot 10^{-3} R^8 {G^{\prime\prime}}^8\Big]^2  \label{eq:LossModOnly}\quad,
\end{align}
taking into account contributions in $\theta_0$ up to second order.\\
In all above equations, numerical coefficients were rounded to two decimal places.
In the Materials and Methods section of this document, we explain how Eqn.~\eqref{eq:SurfTens}, \eqref{eq:ElasMod} and \eqref{eq:LossModOnly} are used to analyse our data.

\section*{\large  Supplementary figures}
\begin{figure*}[h]
\centering
\includegraphics[width=6cm]{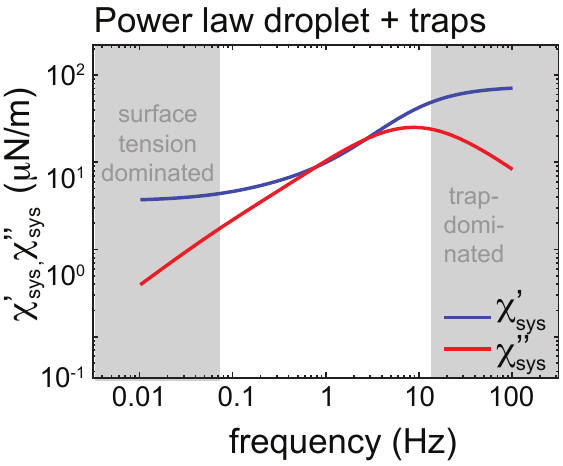}
\caption{ \label{fig:ChiSysPowerLaw}
Effective  spring constant of the combined system of two identical traps in series with a power law droplet (spring constant $k=150\,$pN/$\mu$m).  For low frequencies, the force response of the power law droplet is dominated by surface tension (grey zone, l.h.s.). For high frequencies, the two traps dominate the force response of the model system and the effective spring constant approaches $k/2$ (grey zone, r.h.s.). Outside of this trap-dominated regime, the loss modulus is governed by the power law of the droplet material. 
Parameters are $\theta_0=0.1, R=5\,\mu{\rm m}, \gamma=3\,\mu{\rm N/m}, \eta=0.5\,{\rm Pa s}, A=0.2\,{\rm Pa}, \beta=0.8$. }
\end{figure*}
{\color{black}
\begin{figure*}[h]
\centering
\includegraphics[width=10cm]{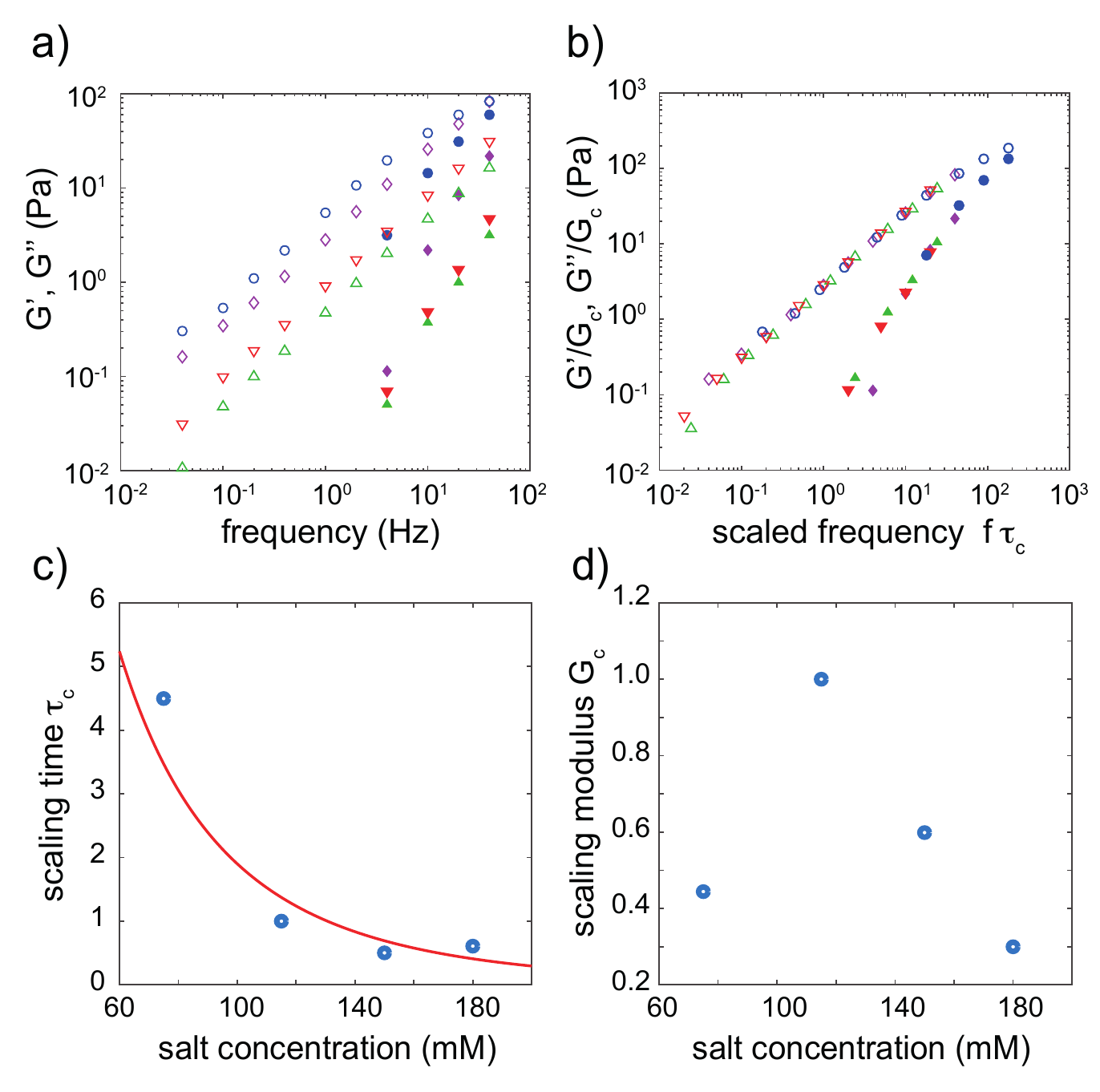}
\caption{ \label{fig:MaxwellPara}
\color{black} 
(a-b) Complex elastic moduli of protein condensates in dependence of salt concentration can be collapsed on top of each other through rescaling of  y-axis and x-axis by a factor $G_c$ and $\tau_c$, respectively. Panel a shows the uncollapsed data, while panel b shows collapsed data after rescaling (blue: 75 mM, purple: 115 mM, red: 150 mM, green: 180 mM, same data as main text). The collapse of rheological data at different salt concentrations through rescaling is analogous to previously reported findings for polyelectrolyte complexes \cite{spru10}.
(c-d) Scaling parameters $\tau_c$ (c) and $G_c$ (d)  at different salt concentrations.  In c, we have fitted a stretched exponential function $A \exp(-B (c/1000)^{1/2} )$ (solid red line). This functional dependence on salt concentration was previously reported for time scales $\tau_c$ for  polyelectrolyte complexes \cite{spru10}. 
We find fit constants of $A=170.4 s$ and $B=14.22\, {\rm M}^{-1/2}$.
 }
\end{figure*}
}

\newpage
\newpage
{ 
\color{black}  {\bf Supplemental Movie 1, caption:} Confocal time-lapse movie  of fluorescent protein droplet (blue, salt concentration $75\,$mM) and adherent beads (yellow). Beads are held by the optical traps. Movement of the mobile trap leads to bead displacement and droplet deformation. The frame interval is $0.5$~s. The pixel size is $200$~nm.
Beads are $2\,\mu$m in size (for illustrative purposes) and the traps are moving apart at a rate of $0.002\,\mu$m/s.
 }
\vspace{1cm}

{ 
\color{black}  {\bf Supplemental Movie 2, caption:} 
Brightfield time-lapse movie of a protein droplet (salt concentration 180 mM) held in place through two adherent beads which are trapped by the two optical traps. The droplet undergoes visible thermal shape fluctuations over time. The frame interval is $0.01$~s. The pixel size is $90$~nm.
 }